\PassOptionsToPackage{table,xcdraw}{xcolor}

\documentclass[conference,a4paper]{APSIPA2021}
\usepackage{array, amsmath, amssymb, amsfonts, bm}
\usepackage{graphicx}
\usepackage{multirow, booktabs}
\usepackage{threeparttable}
\usepackage[backend=biber,style=ieee,]{biblatex}
\usepackage{xintexpr}
\usepackage{flushend}
\usepackage{lipsum}

\usepackage{microtype}

\usepackage{geometry}
\IEEEoverridecommandlockouts

\begin{document}

\setlength{\abovedisplayskip}{8pt}
\setlength{\belowdisplayskip}{8pt}
\setlength{\abovedisplayshortskip}{8pt}
\setlength{\belowdisplayshortskip}{8pt}
\allowdisplaybreaks[4]

\title{
Run-Time Adaptation of Neural Beamforming \\
for Robust Speech Dereverberation and Denoising
}

\author{
\authorblockN{
Yoto Fujita$^{1,2}$,
Aditya Arie Nugraha$^2$,
Diego Di Carlo$^2$, \\
Yoshiaki Bando$^{3,2}$,
Mathieu Fontaine$^{4,2}$,
and Kazuyoshi Yoshii$^{5,2}$
\vspace{\baselineskip}
}

\authorblockA{
$^1$Graduate School of Informatics, Kyoto University, Japan}

\authorblockA{
$^2$Center for Advanced Intelligence Project (AIP), RIKEN, Japan}

\authorblockA{
$^3$AIRC, National Institute of Advanced Industrial Science and Technology (AIST), Japan}

\authorblockA{
$^4$LTCI, T\'{e}l\'{e}com Paris, Institut Polytechnique de Paris, France
}

\authorblockA{
$^5$Graduate School of Engineering, Kyoto University, Japan}
}

\maketitle

\setlength{\abovedisplayskip}{5pt}
\setlength{\belowdisplayskip}{5pt}
\allowdisplaybreaks[4]

\begin{abstract}
This paper describes speech enhancement
 for real-time automatic speech recognition (ASR)
 in real environments.
A standard approach to this task
 is to use neural beamforming 
 that can work efficiently in an online manner.
It estimates the masks of clean dry speech from a noisy echoic mixture spectrogram
 with a deep neural network (DNN)
 and then computes a enhancement filter used for beamforming.
The performance of such a supervised approach, however, 
 is drastically degraded under mismatched conditions.
This calls for run-time adaptation of the DNN.
Although the ground-truth speech spectrogram required for adaptation is not available at run time,
 blind dereverberation and separation methods 
 such as weighted prediction error (WPE) 
 and fast multichannel nonnegative matrix factorization (FastMNMF)
 can be used for generating pseudo ground-truth data from a mixture.
Based on this idea,
 a prior work proposed 
 a dual-process system based on a cascade of WPE and
 minimum variance distortionless response (MVDR) beamforming
 asynchronously fine-tuned by block-online FastMNMF.
To integrate the dereverberation capability 
 into neural beamforming and make it fine-tunable at run time,
 we propose to use weighted power minimization distortionless 
 response (WPD) beamforming,
 a unified version of WPE and minimum power distortionless response (MPDR),
 whose joint dereverberation and denoising filter 
 is estimated using a DNN.
We evaluated the impact of run-time adaptation
 under various conditions
 with different numbers of speakers, 
 reverberation times, 
 and signal-to-noise ratios (SNRs).
 


\end{abstract}


\section{Introduction}
\label{sec:introduction}

Robust speech enhancement is a key technique
 in practical automatic speech recognition (ASR) systems
 that work in real time in real environments.
For this purpose, 
 one may use
 blind source separation (BSS) methods
 such as multichannel nonnegative matrix factorization (MNMF) 
 \cite{ozerov2009multichannel,sawada2013multichannel},
 independent low-rank matrix analysis (ILRMA) 
 \cite{kitamura2016determined}, and
 FastMNMF \cite{ito2019fastmnmf,sekiguchi2020fast}.
Among these,
 FastMNMF is a state-of-the-art method
 that has been shown to outperform MNMF and ILRMA \cite{sekiguchi2020fast}. 
These methods are based on unsupervised learning
 (maximum likelihood estimation)
 of probabilistic models of mixture signals
 and are thus essentially free 
 from the condition mismatch problem of supervised learning methods.
However, 
 these methods are hard to use for real-time systems
 due to the computationally demanding iterative optimization 
 required at run time.

\begin{figure}
    \centering
    \includegraphics[width=.98\hsize]{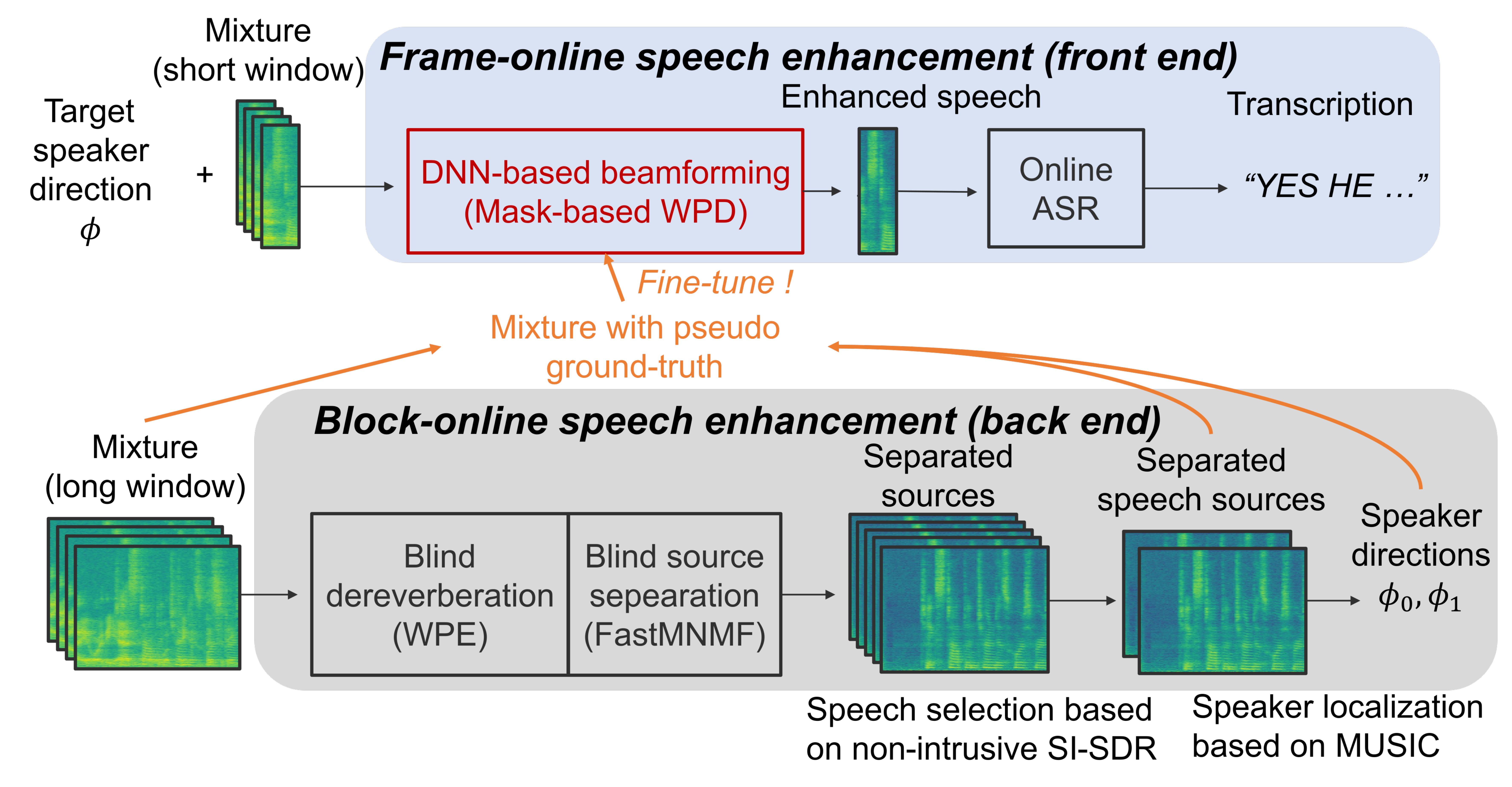}
    \vspace{-2mm}
    \caption{The overview of our proposed  adaptation of neural beamforming based on blind dereverberation by WPE and blind source separation by FastMNMF.}
    \label{fig:overview}
    \vspace{-3mm}
\end{figure}

In recent years, 
 deep neural networks (DNNs) 
 have widely been used for speech enhancement.
This approach, for example, performs
 direct mapping from a noisy mixture 
 into multiple speech sources~\cite{wang2018supervised},
 mixture-conditioned 
 deep speech generation~\cite{pascual2017segan, lu2022conditional}
 for single-channel data,
 and neural beamforming 
 with a DNN-based mask estimator~\cite{heymann2017generic}
 for multichannel data.
In general, 
 while supervised training of a DNN is computationally demanding,
 inference with the DNN works fast
 even on edge devices~\cite{sekiguchi2022direction}.
Considering the low-latency and high-performance nature
 and the potential contribution to the ASR~\cite{chime4}, 
 we focus on DNN-based beamforming (a.k.a. neural beamforming)
 as a front end of real-time distant ASR.

In neural beamforming,
 a DNN is used for estimating speech masks
 in the time-frequency domain.
To separate clean speech from noisy speech mixture,
 the spatial covariance matrices (SCMs) of speech and noise
 are computed from the estimated masks
 and a enhancement filter used for beamforming
 is computed from the SCMs.
The DNN
 is trained in a supervised manner 
 by using pairs of noisy mixture and clean speech, 
 often with the directions of the target speakers
 \cite{sekiguchi2022direction,chen2018multi}.
However, 
 it is not realistic 
 to collect training data
 covering diverse acoustic environments 
 that the model is potentially applied to.
This makes the model less robust 
 to unseen acoustic environments.

A promising solution to this problem
 is a run-time adaptation of neural beamforming~\cite{sekiguchi2022direction, aizawa2023unsupervised}.
The biggest challenge in this task
 is that no ground-truth data 
 (clean speech) are available
 unlike in standard offline benchmarks.
To solve this problem,
 one can fine-tune a DNN-based mask estimator 
 using \textit{pseudo} ground-truth data
 given by FastMNMF.
In a dual-process system~\cite{sekiguchi2022direction},
 a light-weight 
 minimum variance distortionless response (MVDR) beamforming (front end)
 is used for streaming speech enhancement,
 where the mask estimator is fine-tuned 
 with the target speech 
 separated by asynchronously-running FastMNMF (back end),
 often with the direction of the target speaker.
It has been shown that the ASR performance
 tends to improve along with the amount of fine-tuning data
 (e.g., multi-party conversation data)
 \cite{sekiguchi2022direction}.
This system also uses 
 weighted prediction error (WPE)~\cite{nakatani2010speech, yoshioka2012generalization},
 a popular blind dereverberation method,
 before MVDR beamforming and FastMNMF for improved ASR.
However, since the adaptation is only applied 
 to the mask estimator for beamforming, 
the adaptation capability of this system
 is thus limited to MVDR beamforming only.

In this paper, 
 we propose run-time adaptation of neural beamforming
 for joint speech dereverberation and denoising.
Since WPE works stably in various environments 
 thanks to the unsupervised nature~\cite{sekiguchi2022direction},
 we aim to draw its full potential with its neural extension. 
Specifically,
 we use weighted power minimization distortionless response (WPD) beamforming
 \cite{nakatani2019unified, nakatani2019simultaneous}, 
 a unified version of WPE and 
 a minimum power distortionless response (MPDR),
 where a DNN-based mask estimator
 is used to estimate a unified dereverberation and denoising filter.
Both the speech dereverberation and denoising functions of the system
 can be adapted to a test environment
 while considering the mutual dependency of both tasks.
We 
 comprehensively investigate acoustic conditions
 in which the adaptation effectively contributes
 to the improvement of speech enhancement and ASR. 




\section{Related Work}
\label{sec:related_work}

This section reviews
 speech dereverberation based on WPE
 \cite{nakatani2010speech, yoshioka2012generalization},
 speech enhancement based on MPDR beamforming,
 joint dereverberation and denoising
 based on WPD beamforming
 \cite{nakatani2019unified, nakatani2019simultaneous},
 and BSS based on FastMNMF \cite{sekiguchi2020fast}.

\subsection{Dereverberation}

WPE is a well-known blind dereverberation method 
 based on an autoregressive model of late reverberation. 
Let $\mathbf{x}_{ft} \in \mathbb{C}^{M}$ 
 be the short-time Fourier transform (STFT) spectrum 
 of an observed mixture 
 captured by an $M$-channel microphone array
 at frequency $f \in [1,F]$
 and time frame $t \in [1,T]$,
 where
 $F$ is the number of frequency bins and 
 $T$ is the number of frames.
We assume $\mathbf{x}_{ft}$ can be decomposed as follows:
\begin{align}\label{eq:1}
    \mathbf{x}_{ft} 
    = 
    \mathbf{d}_{ft} + \mathbf{r}_{ft},
\end{align}
where 
 $\mathbf{d}_{ft} \in \mathbb{C}^{M}$ 
 is direct signals with early reflections,
 $\mathbf{r}_{ft} \in \mathbb{C}^{M}$ 
 is the spectrum of late reverberation.
%
The late reverberation is assumed 
 to be the weighted sum of past observations as follows:
\begin{align}\label{eq:2}
    \mathbf{r}_{ft} 
    = 
    \sum_{\tau=b}^{L} 
    \mathbf{W}_{f\tau}^\mathsf{H} \mathbf{x}_{f,t-\tau},
\end{align}
where 
 $\mathbf{W}_{f\tau} \in \mathbb{C}^{M \times M}$ 
 is a mixing filter for delay $\tau$,
 $L$ is a tap length,
 and $b$ is a prediction delay
 representing the boundary between
 the early reflections and late reverberation.
The target 
 $\mathbf{d}_{ft}$ 
 is thus given by
\begin{align}\label{eq:3}
     \mathbf{d}_{ft} 
     = 
     \mathbf{x}_{ft} 
     - 
     \sum_{\tau=b}^{L} \mathbf{W}_{f\tau}^\mathsf{H} \mathbf{x}_{f,t-\tau}.
\end{align}

The filter is estimated
 by minimizing the weighted power of the estimated direct signal
 as follows:
\begin{align}\label{eq:4}
    \widehat{\mathbf{W}}_f 
    = 
    \underset
    {\mathbf{W}_f}
    {\operatorname{argmin}} \:
    \mathbb{E}_t \left[
    \frac
    {|\mathbf{x}_{ft}  
    - 
    \sum_{\tau=b}^{L} \mathbf{W}_{f\tau}^\mathsf{H} \mathbf{x}_{f,t-\tau} |^2}
    {\sigma_{ft}^2} \right],
\end{align}
where $\sigma_{ft}^2$ represents 
 the time-varying power spectral density (PSD) 
 of the target speech. 
The PSD can be obtained 
 through iterative updates of the estimated target speech 
 and its power~\cite{drude2018integrating}, 
 or by source mask estimation using a DNN~\cite{togami2020joint}.

\subsection{Speech Enhancement}

The multichannel signal model 
 in \eqref{eq:1} 
 can be rewritten
 by decomposing a target speech signal  
 as the product of a steering vector 
 $\mathbf{a}_f \in \mathbb{C}^M$
 and a source $s_{ft} \in \mathbb{C}$, 
 while considering as noise 
 other components
 including non-target components,
 early reflections, and late reverberations as follows:
\begin{align}\label{eq:5}
    \mathbf{x}_{ft} 
    = 
    \mathbf{a}_f s_{ft} 
    +
    \mathbf{n}_{ft},
\end{align}
where $\mathbf{n}_{ft}$ is the spectrum of noise.
The target speech $\widehat{d}_{ft}$ 
 is estimated by applying an enhancement filter 
 $\mathbf{w}_{f0} \in \mathbb{C}^M$ 
 to the mixture $\mathbf{x}_{ft}$ as follows:
\begin{align}\label{eq:6}
    \widehat{d}_{ft} 
    = 
    \mathbf{w}_{f0}^\mathsf{H} \mathbf{x}_{ft} .
\end{align}


In MPDR beamforming~\cite{vincent2018audio},
 the filter is estimated
 by minimizing the power of 
 the observed mixture 
 $\mathbf{x}_{ft}$, 
 while maintaining a distortionless response 
 in the direction of the steering vector 
 $\mathbf{a}_f$ as follows:
\begin{align}
    \widehat{\mathbf{w}}_{f0}^{\mbox{\tiny{MPDR}}} 
    &= 
    \underset
    {\mathbf{w}_{f0}}
    {\operatorname{argmin}} \:
    \mathbb{E}_t 
    \left[ |\mathbf{w}_{f0}^\mathsf{H} \mathbf{x}_{ft}|^2 \right]
    \quad \mathrm{s.t.} \quad
    \mathbf{w}_{f0}^\mathsf{H} \mathbf{a}_f = 1 \label{eq:mpdr_obj}.
\end{align}
The closed-form solution of the optimal filter
 is given by
\begin{align}
    \widehat{\mathbf{w}}_{f0}^{\mbox{\tiny{MPDR}}} 
    &= 
    \frac
    {\mathbf{K}_{f}^{-1} \mathbf{a}_f}
    {\mathbf{a}^\mathsf{H} \mathbf{K}_{f}^{-1} \mathbf{a}_f},
    \label{eq:MPDR_opt}
\end{align}
where 
 $\mathbf{K}_{f} = \mathbb{E}_t [ \mathbf{x}_{ft}\mathbf{x}_{ft}^\mathsf{H} ]$ 
 is the SCM of the mixture.
 
\subsection{Joint Speech Dereverberation and Denoising}

WPD beamforming is formulated by integrating WPE and MPDR beamforming
 for jointly dereverberation and enhancement.
Specifiacally, using \eqref{eq:3} and \eqref{eq:6},
the signal obtained with WPD beamforming
 is given by
\begin{align}
    \widehat{d}_{ft}
    &= 
    \mathbf{w}_{f0}^\mathsf{H} \left( \mathbf{x}_{t} 
    + 
    \sum_{\tau=b}^{L} \mathbf{W}_{f\tau}^\mathsf{H} \mathbf{x}_{f,t-\tau} \right)
    = 
    \overline{\mathbf{w}}_f^\mathsf{H} \overline{\mathbf{x}}_{ft}, \label{eq:wpd}
\end{align}
where 
 $\overline{\mathbf{w}}_f \in \mathbb{C}^{(L-b+1)M}$ 
 is an integrated filter consisting of  
 $\{\mathbf{w}_{ft}\}_{t=0,b,\dots,L}$
 and 
 $\overline{\mathbf{x}}_{ft} \in \mathbb{C}^{(L-b+1)M}$ 
 is the concatenation of the current and past observations
 $\{\mathbf{x}_{f,t-\tau}\}_{\tau=0,b,\dots,L}$.
 
Using \eqref{eq:4} and \eqref{eq:mpdr_obj}, 
 the filter
 $\overline{\mathbf{w}}_f$ 
 is estimated as
\begin{align}\label{eq:15}
    \widehat{{\mathbf{w}}}_f^{\mbox{\tiny{WPD}}}
    = 
    \underset
    {\overline{\mathbf{w}}_f}
    {\operatorname{argmin}} \:
    \mathbb{E}_t \left[
    \frac
    {|\overline{\mathbf{w}}_f^\mathsf{H} \overline{\mathbf{x}}_{ft}|^2}
    {\sigma_{ft}^2} \right] 
    \:\: \mathrm{s.t.} \:\: 
    \mathbf{w}_{f0}^\mathsf{H} \mathbf{a}_f = 1.
\end{align}
The closed-form solution of the optimal filter is given by
\begin{align}\label{eq:16}
    \widehat{{\mathbf{w}}}_{f}^{\mbox{\tiny{WPD}}} 
    = 
    \frac
    {\overline{\mathbf{\mathbf{K}}}_f^{-1} \overline{\mathbf{a}}_f}
    {\overline{\mathbf{a}}_f^\mathsf{H} \overline{\mathbf{\mathbf{K}}}_f^{-1} \overline{\mathbf{a}}}_f,
\end{align}
where 
 $\overline{\mathbf{K}}_f = \mathbb{E}_t [ {\overline{\mathbf{x}}_{ft}\overline{\mathbf{x}}_{ft}^\mathsf{H}}{\sigma_{ft}^{-2}} ]$ 
 is the SCM of the mixture
 compensated by the PSD of the target speech, 
 and 
 $\overline{\mathbf{a}}_f \in \mathbb{C}^{(L-b+1)M}$ 
 is the concatenation of the steering vector 
 $\mathbf{a}_f$ 
 and a zero vector 
 $\mathbf{0} \in \mathbb{R}^{(L-b)M}$.

Using the SCM of the target speech
 $\overline{\mathbf{R}}_{f} = \overline{\mathbf{a}}_f \overline{\mathbf{a}}_f^\mathsf{H} |s_{ft}|^2$, 
 \eqref{eq:16} can be reformulated as:
\begin{align}\label{eq:17}
    \widehat{{\mathbf{w}}}_f^{\mbox{\tiny{WPD}}}
    = 
    \frac
    {\overline{\mathbf{K}}_f^{-1}\overline{\mathbf{R}}_{f}}
    {\mathrm{tr}(\overline{\mathbf{K}}_f^{-1}\overline{\mathbf{R}}_{f})}
    \overline{\mathbf{u}}_q,
\end{align}
where 
 $\overline{\mathbf{u}}_q \in \mathbb{R}^{(L-b+1)M}$ 
 is a one-hot vector
 whose $q$-th element (reference channel) takes one
 and zero otherwise.

In mask-based WPD beamforming, 
 the PSD $\sigma_{ft}^2$ 
 and SCM $\overline{\mathbf{R}}_{f}$ of the target speech
 are computed with 
 $\widehat{\mathbf{s}}_{ft} = \omega_{ft} \mathbf{x}_{ft}$, 
 where a time-frequency (TF) speech mask 
 $\omega_{ft} \in [0, 1]$ 
 is estimated by a DNN
 \cite{nakatani2020dnn,zhang2020end}.

\subsection{Blind Source Separation}

%

The general goal of BSS is to separate a mixture spectrogram
 $\{\mathbf{x}_{ft}\}_{f,t=1}^{F,T}$
 into $N$ source spectrograms 
 $\{\{\mathbf{x}_{nft}\}_{f,t=1}^{F,T}\}_{n=1}^N$,
 where $\mathbf{x}_{ft}, \mathbf{x}_{nft} \in \mathbb{C}^M$.
In modern BSS methods,
 each source $\mathbf{x}_{nft}$ is typically assumed 
 to follow an $M$-variate circularly-symmetric 
 complex Gaussian distribution as follows:
\begin{align}\label{eq:20}
    \mathbf{x}_{nft}
    \sim 
    \mathcal{N}_\mathbb{C} (\mathbf{0}, \lambda_{nft} \mathbf{G}_{nf}),
\end{align}
where $\lambda_{nft}$ and $\mathbf{G}_{nft}$ 
 are the PSD and SCM of the source $n$.
Assuming the additivity of complex spectrograms
 and using the reproductive property of the Gaussian distribution,
 the mixture $\mathbf{x}_{ft}$ is given by 
\begin{align}
    \mathbf{x}_{ft}
    \sim 
    \mathcal{N}_\mathbb{C} 
    \left(\mathbf{0}, \sum_{n=1}^N 
    \lambda_{nft} \mathbf{G}_{nf}\right).
    \label{eq:bss_lk}
\end{align} 
 
In MNMF~\cite{ozerov2009multichannel,sawada2013multichannel} 
 and its variants including FastMNMF~\cite{sekiguchi2020fast},
 the PSDs $\{\lambda_{nft}\}_{f,t=1}^{F,T}$ are 
 factorized with NMF as follows:
\begin{align}
 \lambda_{nft} = \sum_{k=1}^K u_{nkf} v_{nkt},
\end{align}
where $\mathbf{u}_{nk} \in \mathbb{R}_+^F$ 
 is a basis vector,
 $\mathbf{v}_{nk} \in \mathbb{R}_+^T$
 is an activation vector,
 and $K$ is the number of bases.
In FastMNMF,
 the SCM $\mathbf{G}_{nf}$ 
 is also factorized as follows:
\begin{align}
 \mathbf{G}_{nf} = \mathbf{Q}_f^{-1} \mathrm{Diag}(\tilde{\mathbf{g}}_n) \mathbf{Q}_f^{-\mathsf{H}},
\end{align}
where $\mathbf{Q}_f \in \mathbb{C}^{M \times M}$
 is a time-invariant diagonalization matrix 
 and $\tilde{\mathbf{g}}_n \in \mathbb{R}_+^M$
 is a frequency-invariant nonnegative vector.

The model parameters 
 $\{\mathbf{u}_{nk}\}_{n,k=1}^{N,K}$,
 $\{\mathbf{v}_{nk}\}_{n,k=1}^{N,K}$,
 $\{\mathbf{Q}_f\}_{f=1}^{F}$,
 and $\{\tilde{\mathbf{g}}_n\}_{n=1}^{N}$
 are estimated with an iterative optimization algorithm 
 such that the likelihood of the parameters for the mixture
 given by \eqref{eq:bss_lk} is maximized
 \cite{sekiguchi2020fast}.
Given the optimal parameters,
 the separation filter $\mathbf{w}_{nft}^{\mbox{\tiny{BSS}}}$
 is given by
\begin{align}\label{eq:21}
\widehat{\mathbf{w}}_{nft}^{\mbox{\tiny{BSS}}} 
    = 
    \mathbf{Q}_f^\mathsf{H} \mathrm{Diag} \left( \frac
    {\lambda_{nft}\tilde{\mathbf{g}}_n}
    {\sum_{n'} \lambda_{n'ft}\tilde{\mathbf{g}}_{n'}}
    \right) \mathbf{Q}_f^{-\mathsf{H}}
    \mathbf{u}_q ,
\end{align}
where $\mathbf{u}_q \in \{0,1\}^M$ 
 is a one-hot vector
 whose $q$-th element (reference channel) takes one
 and zero otherwise.


\section{Proposed method}
\label{sec:proposed_method}

This section describes 
 the proposed adaptive joint dereverberation and denoising system
 based on a dual-process architecture
 consisting of mask-based WPD beamforming with FastMNMF-guided fine-tuning.
This system uses the WPD beamforming 
 to perform low-latency speech dereverberation and denoising,
 resulting in a single-channel speech signal useful for the ASR system.
To be adaptive to dynamic environments,
 its DNN-based mask estimator is fine-tuned at run time
 using speech signals dereverberated and separated 
 with high-latency yet environment-robust WPE and FastMNMF.

\subsection{Joint Neural Speech Dereverberation and Denoising}

Given a mixture spectrogram 
 $\mathbf{X} \triangleq 
 \{\mathbf{x}_{ft} \in \mathbb{C}^M\}_{f=1,t=1}^{F,T}$
 with target speaker DOAs
 $\bm\phi \triangleq 
 \{\phi_t \in [0, 2\pi]\}_{t=1}^T$, 
 a DNN $\mathcal{F}_{\mathbf{\Theta}}$ 
 parameterized by $\mathbf{\Theta}$ 
 is used for estimating TF masks 
 $\bm\omega \triangleq \{\omega_{ft}\}_{f=1,t=1}^{F,T}$ 
 as follows:
\begin{align}
    \bm\omega
    =
    \mathcal{F}_{\mathbf{\Theta}}
    \left(\mathbf{X}, \bm\phi\right).
\end{align}
%

The PSD $\widehat\sigma_{ft}^2$ and 
 SCM $\widehat{\mathbf{R}}_{f}$ 
 of the target speech 
 can be computed using the mask estimate as follows:
\begin{align}
    \widehat\sigma_{ft}^2 
    &= 
    \frac{1}{M} \sum_{m=1}^M |\omega_{ft}x_{ftm}|^2 , \\
    \widehat{\mathbf{R}}_{f} 
    &=
    \frac{1}{T} \sum_{t=1}^T \widehat{{\mathbf{s}}}_{ft} \widehat{{\mathbf{s}}}_{ft},
\end{align}
where
$\widehat{{\mathbf{s}}}_{ft} = [\omega_{ft}\mathbf{x}_{ft}^\top, \omega_{f,t-b}\mathbf{x}_{f,t-b}^\top,\dots,\omega_{f,t-L}\mathbf{x}_{f,t-L}^\top]^\top \in \mathbb{C}^{(L-b+1)M}$.
The WPD filter 
 $\widehat{{\mathbf{w}}}_f^{\mbox{\tiny{WPD}}}$ 
 is then computed from
 $\widehat{\mathbf{\mathbf{K}}}_f 
 = \sum_t
 \overline{\mathbf{x}}_{ft}
 \overline{\mathbf{x}}_{ft}^\mathsf{H}
 \widehat{\sigma}_{ft}^{-2}$
 and $\widehat{\mathbf{R}}_{f}$
 as in \eqref{eq:17}.
Finally, 
 the target speech signal 
 $\widehat{d}_{ft}$ 
 corresponding to the DOA $\phi_t$ 
 is obtained by applying
 $\widehat{{\mathbf{w}}}_f^{\mbox{\tiny{WPD}}}$ 
 to 
 $\overline{\mathbf{x}}_{ft}$ as in \eqref{eq:wpd}.

\subsection{Pretraining of Mask Estimator}
\label{sec:pretraining_mask}

The DNN-based mask estimator 
 $\mathcal{F}_{\mathbf{\Theta}}$ 
 is pretrained using triples
 consisting of an $M$-channel mixture, 
 a reference speech, 
 and a target DOA. 
It is optimized to minimize
 the negative signal-to-distortion ratio (SDR) 
 between the estimated time-domain speech signal 
 $\widehat{\mathbf{d}} \in \mathbb{R}^S$ 
 and the reference time-domain speech signal 
 $\mathbf{d}^{\mbox{\scriptsize{ref}}} \in \mathbb{R}^S$ given by
\begin{align}\label{eq:25}
    \mathcal{L}
    =
    - 10 \, \log_{10} \frac
    {\widehat{\mathbf{d}}^{\top} \widehat{\mathbf{d}}}
    {(\widehat{\mathbf{d}} - \mathbf{d}^{\mbox{\scriptsize{ref}}})^\top (\widehat{\mathbf{d}} - \mathbf{d}^{\mbox{\scriptsize{ref}}})},
\end{align}
where $S$ denotes the number of samples. 
The estimated time-domain speech signal
 $\widehat{\mathbf{d}}$ 
 is obtained by applying the inverse STFT 
 to the estimated speech signal 
 in the STFT domain 
 $\{\widehat{d}_{ft}\}_{f=1,t=1}^{F,T}$.
 
\subsection{Run-Time Adaptation of Mask Estimator}

To make the mask estimator $\mathcal{F}_{\mathbf{\Theta}}$ 
 adaptive to environmental changes,
 we fine-tune it at run time
 using triples of the observed mixture 
 $\{\mathbf{x}_{ft}\}_{f=1,t=1}^{F',T'}$, 
 the pseudo ground-truth speech signal $\mathbf{d}^{\mbox{\scriptsize{ref}}}_{p} \in \mathbb{R}^S$, 
 and the pseudo ground-truth DOA $\phi_p$
 to minimize the negative SDR loss in \eqref{eq:25},
 where $F'$ and $T'$ are 
 the number of frequency bins 
 and that of frames 
 of the mixture used for fine-tuning, respectively.

To obtain the pseudo ground-truth speech signal,
 we first dereverberate the mixture 
 $\mathbf{x}_{ft}$ 
 using WPE as follows:
\begin{align}
     \mathbf{x}_{ft}^{\mbox{\scriptsize{dry}}} 
     = 
     \mathbf{x}_{ft} 
     - 
     \sum_{\tau=b}^{L} \mathbf{W}_{f\tau}^\mathsf{H} \mathbf{x}_{f,t-\tau}.
\end{align}
where 
 $\mathbf{x}_{ft}^{\mbox{\scriptsize{dry}}}$
 is the dereverberated mixture 
 and 
 $\mathbf{W}_{f\tau}$
 is a filter for delay $\tau$.
The filter is estimated as \eqref{eq:4}
 with the PSD 
 $\sigma_{ft}^2$ 
 obtained through iterative updates of 
 the estimated dereverberated signal 
 $\mathbf{x}_{ft}^{\mbox{\scriptsize{dry}}}$
 and its power 
 $\sigma_{ft}^2$
 \cite{drude2018integrating}.
Then, we separate
 the sources $\{\{x_{nft}\}_{f,t=1}^{F',T'}\}_{n=1}^N$ 
 from the dereverberated mixture
 using FastMNMF as follows:
\begin{align}
    x_{nft} 
    = 
    \left( \widehat{\mathbf{w}}_{nft}^{\mbox{\tiny{BSS}}} \right)^\mathsf{H} \mathbf{x}_{ft}^{\mbox{\scriptsize{dry}}},
\end{align}
where the separation filter
 $\mathbf{w}_{nft}^{\mbox{\tiny{BSS}}}$  
 is calculated as in \eqref{eq:21}.
We measure the signal quality of each separated source signal
using the reference-less non-intrusive scale-invariant SDR~\cite{leroux2019sdr,kumar2023squim}.
We take $N' (\leq N)$ separated signals
that satisfy a predefined threshold $\alpha$
and consider these as pseudo ground-truth speech signals $\{\mathbf{d}^{\mbox{\scriptsize{ref}}}_{p,n}\}_{n=1}^{N'}$.
Finally, 
 the corresponding $N'$ pseudo ground-truth DOAs 
 $\{\phi_{p,n}\}_{n=1}^{N'}$ 
 are estimated 
 based on multiple signal classification (MUSIC)~\cite{schmidt1986multiple}. 

\begin{figure*}
    \centering
    \includegraphics[width=1.0\hsize]{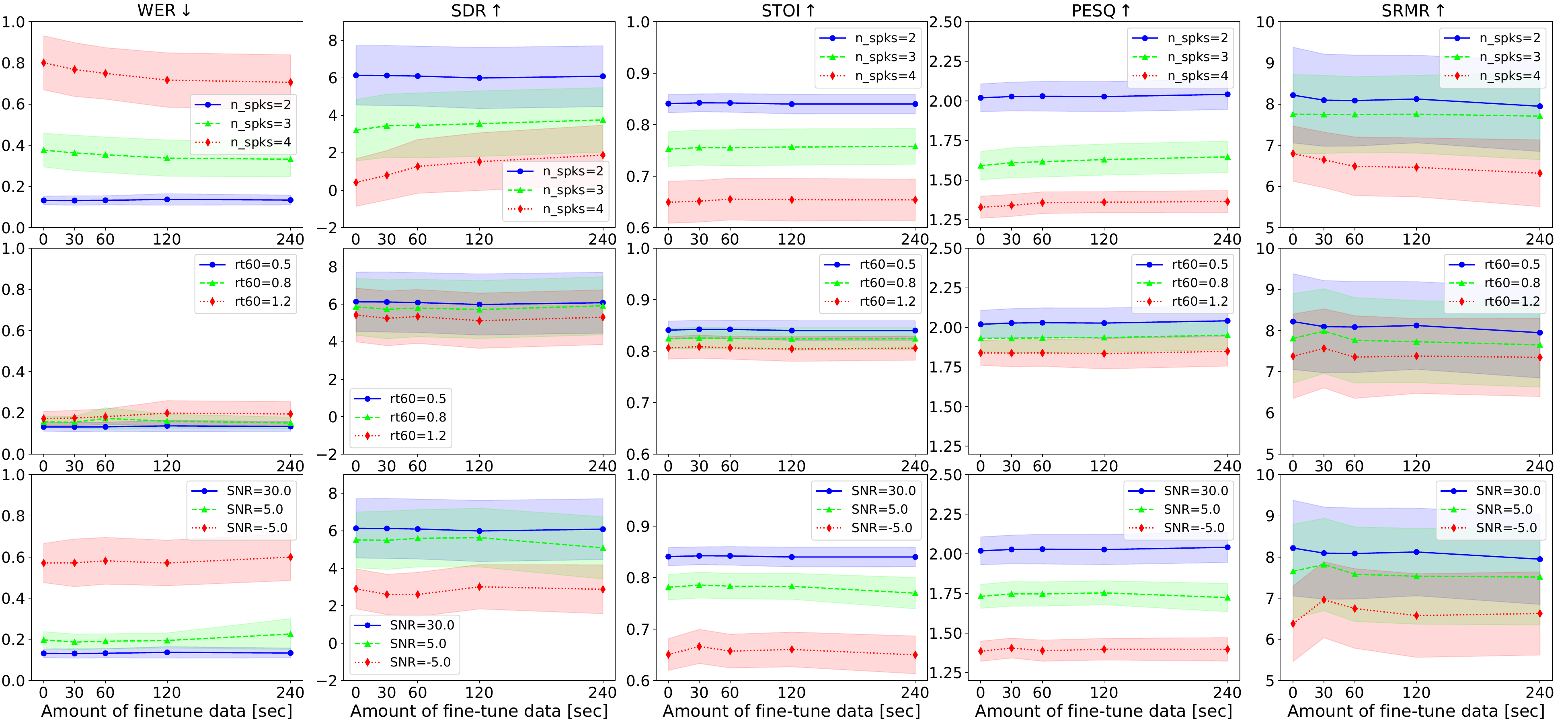}
    \caption{The evaluation result of the joint adaptation of dereverberation and denoising by mask-based WPD beamforming based on blind dereverberation by WPE and BSS by FastMNMF. Blue lines refer to the ``default'' setting, constant across different plots. The shaded regions surrounding each line indicate the 95\% confidence intervals.}
    \label{fig:result}
\end{figure*}

\section{Evaluation}

We report a comprehensive evaluation 
 conducted for assessing the performance of our adaptive system
 in various simulated acoustic environments. 

\subsection{Dataset}

For the mask estimator 
 $\mathcal{F}_{\mathbf{\Theta}}$, 
 we made a \textit{training dataset} comprising 36,000 triples
 and a \textit{validation dataset} comprising 3,600 triples.
Each triple consisted of a 2-second 7-channel mixture signal, 
 a 2-second 7-channel target speech signal, 
 and the corresponding target DOA.
Each mixture was composed of two speech signals by different speakers, 
 who were stationary during the recording,
 and a diffuse noise signal.
The speech signals were randomly taken 
 from the training set (for the \textit{training dataset}) 
 and the development set (for the \textit{validation dataset})
 of Librispeech~\cite{panayotov15librispeech}, 
 while the diffuse or moving noise signals were randomly taken 
 from the DEMAND dataset~\cite{thiemann2013demand}. 

Both mixture and target speech signals
 were simulated using Pyroomacoustics~\cite{scheibler2018pyroomacoustics}
 by considering a 7-channel circular microphone array, 
 configured to match the geometry constraints of the DEMAND dataset,
 within a 2-dimensional room. 
The room length and width were randomly sampled from uniform distributions
 $\mathcal{U} (7.6 \text{m}, 8.4 \text{m})$ and $\mathcal{U} (5.6 \text{m}, 6.4 \text{m})$, respectively.
The 2-dimensional coordinates of the array center were sampled from
 $\mathcal{U} (3.6 \text{m}, 4.4 \text{m})$
 and $\mathcal{U} (2.6 \text{m}, 3.4 \text{m})$, respectively.
The distance between the array center and each speaker 
 was sampled from $\mathcal{U} (1 \text{m}, 2 \text{m})$.
The reverberation time (RT60) for the mixtures was varied 
 between 0.25 and 0.7 seconds.
The target speech signals,
 which were supposed to include the direct path and early component,
 were simulated in the same rooms as the mixtures, 
 but with the RT60 fixed at 0.25 seconds. 
Indoor noise signals from the DEMAND dataset, 
 excluding the environment ``PSTATION'' that was used for the test set (see below), 
 were randomly selected and added to the simulated mixtures, 
 with an SNR between -5.0 and 5.0 dB.


To evaluate the effectiveness of our adaptation method 
 in various acoustic environments, 
 we prepared a \textit{test dataset} consisting of seven distinct simulation settings,
 i.e., one ``default'' setting and six variations.
The default setting considered mixtures of two speakers
 with an RT60 of 0.5 seconds, an SNR of 30.0 dB,
 and other parameters were randomly sampled as in the pretraining dataset. 
For the other six settings, 
 we independently varied the number of stationary speakers (3 and 4), 
 the RT60s (0.8 and 1.2 seconds), 
 and the SNR (5.0 and -5.0 dB).
Speech signals were taken from the test set of Librispeech, 
 with diffuse or moving noise signals from the environment ``PSTATION'' 
 of the DEMAND dataset. 
For each setting,
 we generated 30 triples, 
 each consisting of an approximately 8-minute mixture, 
 the target speech signal, 
 the target DOA, 
 and corresponding transcriptions. 
The first four minutes of each recording 
 were used for fine-tuning the mask estimator, 
 with the remaining duration reserved for evaluation.

\subsection{Experimental Settings}

Both the front and back ends 
 operated in the STFT domain.
The STFT coefficients were computed using
 a window size of 1024 ($F=513$) 
 with a hop length of 256.

For the mask-based WPD beamforming, 
 the prediction delay and the tap size 
 of the convolutional filter 
 were set to $b=3$ and $L=8$, respectively.
The TF mask was estimated using a DNN as in \cite{sekiguchi2022direction}.
The DNN was composed of
 a preprocessing network, a direction attractor network (DAN),
 and a bidirectional long short-term memory (BLSTM) network.
The preprocessing network took as input
 the concatenation 
 of the log magnitude of the mixture, 
 the inter-channel phase difference, 
 and the beamforming output
 by delay-sum beamforming,
 while the DAN took the target DOA as input.
Given the outputs of these two networks,  
 the BLSTM then estimates a TF mask.
This mask estimator was pretrained on the \textit{training dataset}
 using the AdamW optimizer with a learning rate of $10^{-4}$
 and a batch size of 4.
The model was trained for 20 epochs, 
 and the model with the lowest validation negative SDR loss
 was selected for evaluation.

For the joint adaptation of mask-based WPD beamforming, 
 the prediction delay, 
 the tap size, 
 and the number of iterations for parameter updates in WPE
 were set to $b=3$ and $L=13$, and $3$, 
 respectively.
The number of sources, 
 the number of bases, 
 and the number of iterations for parameter updates in FastMNMF
 were set to $N=5$, $K=16$, and $200$, respectively. 
The threshold for non-intrusive SI-SDR 
 was set to $\alpha=10.0$.
The window size of the mixture 
 given as the input to FastMNMF 
 was set to 30 seconds.
The learning rate was set to $4\times10^{-5}$, 
 and the batch size was 4 for fine-tuning.
To stabilize the fine-tuning, 
 we added the same amount of the pretraining data 
 to the fine-tuning data
 so a batch may contain these two types of data.
 
The evaluation compared different amounts 
 of fine-tuning data (30, 60, 120, and 240 seconds) 
 against several key metrics. 
These metrics include word error rate (WER), 
 signal-to-distortion ratio (SDR), 
 short-time objective intelligibility (STOI), 
 perceptual evaluation of speech quality (PESQ), 
 and speech-to-reverberation modulation energy ratio (SRMR). 
We used a Transformer-based encoder-decoder ASR model
 from the SpeechBrain toolkit~\cite{ravanelli2021speechbrain}
 to measure WER.
This ASR model was trained on the Librispeech dataset.
For all metrics except WER, 
 higher values are better.



\subsection{Experimental Results}

Figure \ref{fig:result} shows the evaluation results 
 of the joint adaptation of dereverberation and denoising 
 using mask-based WPD beamforming
 with different durations of fine-tuning data 
 and various simulation settings. 
The upper left plot in the figure 
 indicates that our adaptation method 
 improved WER, SDR, STOI, and PESQ
 across different numbers of stationary speakers,
 which proves the effectiveness of our adaptation.
These improvements seem to benefit from 
 the robust separation capability of FastMNMF
 for stationary sources.

However, 
 when we used a large amount of fine-tuning data,
 the WER was slightly degraded
 as the RT60 increased or the SNR decreased.
This would be because 
 the pretrained mask-based WPD beamforming already has 
 a strong capability of dereverberation, 
 and FastMNMF is less robust to a mixture 
 with moving sources. 
In contrast, 
 when we used a small amount of fine-tuning data,
 the ASR performance hardly changed
 and the other metrics, 
 STOI, PESQ, and SRMR 
 improved in noisy or reverberant conditions.
This suggests that 
 30 seconds of fine-tuning data
 is optimal for practical use.

%
%




\section{Conclusion}

This paper proposes a run-time adaptation method
 for the joint neural dereverberation and denoising with mask-based WPD beamforming
 using fine-tuning data obtained 
 using WPE and FastMNMF.
Evaluations showed robust improvements
 of the ASR performance
 across different numbers of stationary speakers,
 RT60s, and SNRs when we used a small amount of fine-tuning data.
For future work, 
 BSS methods that are capable of dealing with 
 more various acoustic conditions 
 should be investigated 
 to improve the ASR performance
 even in noisy conditions with moving sources.

\section*{Acknowledgment}

This work was partially supported 
by JST FOREST No. JPMJFR2270, 
JSPS KAKENHI Nos. 24H00742, 23K16912, and 23K16913,
ANR Project SAROUMANE (ANR-22-CE23-0011),
and Hi!~Paris Project MASTER-AI. 

\begingroup
\newcommand{\myfontsize}{\fontsize{10.}{11.}\selectfont}
\def\baselinestretch{1.02}\let\normalsize\myfontsize\normalsize
\printbibliography
\endgroup

\end{document}